\newcommand{\be}{\begin{equation}}
\newcommand{\ee}{\end{equation}}
\newcommand{\brr}{\begin{eqnarray}}
\newcommand{\err}{\end{eqnarray}}
\newcommand{\nn}{\nonumber}
\newcommand{\bd}{\begin{displaymath}}
\newcommand{\ed}{\end{displaymath}}

\newcommand{\bib}{\bibitem}
\newcommand{\bfig}{\begin{figure}}
\newcommand{\efig}{\end{figure}}
\def\alf{\alpha}

\def\th{\theta}

\def\rpar{\right)}
\def\lpar{\left(}
\def\rbk{\right]}
\def\lbk{\left[}
\def\rbr{\right\}}
\def\lbr{\left\{}
\def\lb{\label}
\def\fka{\mathfrak{a}}
\def\ro{\mbox{\boldmath $\rho$}}

\def\inb{\mbox{\tiny {\rm B}}}
\def\inf{\mbox{\tiny {\rm F}}}

\def\re{\mbox{${\rm Re}$}}
\def\rg{\rangle}
\def\lg{\langle}

\def\hp{\hspace{\parindent}}
\documentclass{iopart}
\usepackage{iopams}
\usepackage{graphicx}
\usepackage[all]{xy}
\begin{document}
\title[Quasiprobability distribution functions for periodic phase-spaces]{Quasiprobability distribution functions for periodic
phase-spaces: I. Theoretical Aspects}
\author{M. Ruzzi, M. A. Marchiolli, E. C. da Silva and D. Galetti}
\address{Instituto de F\'{\i}sica Te\'{o}rica, Universidade Estadual Paulista,
         Rua Pamplona 145, 01405-900, S\~{a}o Paulo, SP, Brazil}
\ead{mruzzi@ift.unesp.br}
\begin{abstract}
An approach featuring $s$-parametrized quasiprobability distribution functions is developed for situations where a circular topology is
observed. For such an approach, a suitable set of angle-angular momentum coherent states must be constructed in appropriate fashion.
\end{abstract}

\section{Introduction}

\hp The importance of phase-space quasiprobability distribution functions in the description of different physical systems can hardly
be overestimated. Apart from its own theoretical interest \cite{r1}, they play a key role in quantum optics \cite{r2},
give an appropriate approach to decoherence \cite{r3}, insights on semiclassical methods and alternative approaches for dynamics of
quantum systems \cite{r4}.

Quasiprobability distribution functions are defined on quantum
phase-spaces. A quantum phase-space formalism generally is based on
a mapping scheme which enables one to relate operators and functions
defined on such phase space \cite{r5,r6}. In this kind of approach,
the quasiprobability distributions are the functions associated with
the density operator. Meanwhile, the plethora of different physical
problems might call for different topologies of the phase space
itself, on which those functions are defined. However, almost all
theoretical techniques regarding quantum phase-spaces are suited for
the all important case of Cartesian position and momentum variables,
where the quantum phase-space domain, spanned by the eigenvalues of
position and linear momentum operators, coincides with the classical
phase-space, a merry coincidence from which much physical knowledge
has been inferred. In other scenarios, the situation is somewhat
different: for systems where a circular topology is required, the
only kind of phase-space quasiprobability distribution function
available in literature is the Wigner function, as proposed by
Mukunda \cite{r7} and subsequently developed and studied in
\cite{r8,r9}. It is worth mentioning that the coherent states are
scarcely studied in this scenario (in comparison with the Cartesian
case) and, as far as the authors knowledge goes, the following are
the main references on the subject
\cite{r10,r11,r12,r13,r14,r15,r16}.

Returning to the Cartesian realm, the Cahill-Glauber (CG) approach provided an unified and meaningful view of different quasiprobability
distribution functions \cite{r17}. In addition, this approach is related to particular orderings of operators in a bosonic expansion,
paving the way for a better understanding of these functions in the context of quantum optics. However, it is unable to deal with other
topologies. In fact, there are no attempts in the literature to deal with the angle-angular momentum phase-space (in the sense discussed
above) and, consequently, to properly define the appropriate quasiprobability distribution functions.

The aim of this paper is to fill this breach by constructing suitable coherent states in a circle topology which are physically
meaningful and closely related with that ones previously introduced in \cite{r10,r13,r14,r15,r16}. Basically, the present approach
consists in establishing a specific set of algebraic properties that leads us to formally characterize the angle-angular momentum
coherent states. The mapping kernel and the associated quasiprobability distribution functions are then properly defined, taking
advantage of this algebraic approach and embodying the desired properties of the CG formalism.

This paper is organized as follows. In section 2 we establish the main properties of the angle-angular momentum coherent states which
allow us to characterize a quantum phase-space with nontrivial topology. Following, in section 3 we define a generalized probability
distribution function through a mapping kernel labeled by elements of this phase space where, in particular, the Husimi, Wigner, and
Glauber-Sudarshan functions are promptly obtained. Moreover, we also derive a hierarchical order among them that consists of a smoothing
process described by a well-defined function for the angle-angular momentum variables. Finally, section 4 contains our conclusions and, finally, the appendix presents some basic results on the quantum mechanics of the angle-angular momentum pair, supporting the results of section 2.

\section{Algebraic properties of the angle-angular momentum coherent states}

To construct the angle-angular momentum coherent states we firstly
introduce a normalized reference state (or vacuum state) through a
continuous superposition of angle eigenstates (which are discussed in the appendix, along with other pertinent details), namely
\be
\lb{dgs}
|0,0 \rg_{\inb} \equiv \int_{- \pi}^{\pi} d \th \, \mathfrak{F}_{\inb}(\th) | \th \rg \; ,
\ee
with complex coefficients
\bd
\mathfrak{F}_{\inb}(\th) \equiv \lg \th | 0,0 \rg_{\inb} = \frac{1}{\sqrt{2 \pi}} \displaystyle\frac{\vartheta_{3} \lpar \left.
\frac{\th}{2} \right| \rmi \fka \rpar}{\sqrt{\vartheta_{3} \lpar 0 | 2 \rmi \fka \rpar}} \qquad \mbox{(boson case)}
\ed
evaluated in terms of the Jacobi theta functions \cite{r19} for $\fka = (2 \pi)^{-1}$. The Jacobi 
$\vartheta_{3}$-function itself reads as
\be
\vartheta_{3} \lpar \left.
z \right| \tau \rpar = \sum_{l=-\infty}^{\infty} \exp \left[ i \pi \tau l^{2}\right] \exp\left[2 \rmi l z \right]. 
\ee
The Jacobi $\vartheta_{3}$-function can be obtained by the Poisson sum applied to the gaussian function \cite{bell}. Note that the second argument of the
$\vartheta_{3}$-function controls its width, and with the value here chosen the normalized vacuum state coincides with the one proposed
in \cite{r10,r13}. Now, when $m$ assumes half-integer values, the complex coefficients $\mathfrak{F}_{\inf}(\th) \equiv
\lg \th | 0,0 \rg_{\inf}$ must be written as follows:
\bd
\mathfrak{F}_{\inf}(\th) = \frac{1}{\sqrt{2 \pi}} \displaystyle\frac{\vartheta_{2} \lpar \left. \frac{\th}{2} \right| \rmi \fka
\rpar}{\sqrt{\vartheta_{2} \lpar 0 | 2 \rmi \fka \rpar}} \qquad \mbox{(fermion case)} \;,
\ed
and the $\vartheta_{2}$-function by its turn is 
\be
\vartheta_{2} \lpar \left.
z \right| \tau \rpar = \sum_{l=-\infty}^{\infty} \exp \left[ i \pi \tau \left(l+\frac{1}{2}\right)^{2}\right] \exp\left[2 \rmi \left(l+\frac{1}{2}\right) z \right]. 
\ee For convenience, we will particularize our results for $m \in \mathbb{Z}$ (boson case) throughout this work.

The next step is to adopt Klauder's prescription for coherent states \cite{r20} through the use of unitary displacement operators, i.e.,
\be
\lb{dcs}
| m, \th \rg \equiv {\bf D}(m,\th) |0,0 \rg
\ee
where
\be
\lb{gen}
{\bf D}(m,\th) = \exp \lpar - \frac{\rmi}{2} \, m \th \rpar \exp ( \rmi m \mathbf{\Theta} ) \exp ( - \rmi \th {\bf J} )
\ee
It is always pertinent to remember that, as it is constructed, the angular unitary displacement operator obeys the periodicity required for angular eigenstates. It is also worth mentioning some basic properties of the ${\bf D}(m,\th)$ displacement operators, which follow from appropriate use of Eq.(\ref{weyl}):
\brr
&\mbox{(i)}& {\bf D}^{\dagger}(m,\th) = {\bf D}(-m,-\th) \; , \nn \\
&\mbox{(ii)}& \Tr \lbk {\bf D}^{\dagger}(m^{\prime},\th^{\prime}) {\bf D}(m,\th) \rbk = \delta_{m^{\prime},m} \,
\delta (\th^{\prime} - \th) \; , \nn
\err
the second property being obtained by means of the multiplication law
\bd
{\bf D}(m^{\prime},\th^{\prime}) {\bf D}(m,\th) = \exp \lbk \frac{\rmi}{2}  \lpar m^{\prime} \th - m \th^{\prime} \rpar \rbk
{\bf D} (m+m^{\prime},\th+\th^{\prime}) \; .
\ed
In addition, the set of coherent states $\{ | m,\th \rg \}$ satisfy two important properties associated with the completeness
relation and the scalar product of two angle-angular momentum coherent states -- namely,
\bd
\mbox{(iii)} \sum_{m \in \mathbb{Z}} \int_{-\pi}^{\pi} \frac{d \th}{2\pi} \, | m,\th \rg \lg m,\th | = {\bf 1}
\ed
and
\brr
\fl \mbox{(iv)} \; \; \lg m^{\prime},\th^{\prime} | m,\th \rg &=& \exp \lbr - \frac{1}{2} (m-m^{\prime})^{2} + \frac{\rmi}{2} \lbk
(m \th^{\prime} - m^{\prime} \th) + (m-m^{\prime})(\th - \th^{\prime}) \rbk \rbr \nn \\
\fl & & \times \displaystyle\frac{\vartheta_{3} \lpar \left. \frac{1}{2} (\th - \th^{\prime}) + \frac{\rmi}{2} (m-m^{\prime})
\right| 2 \rmi \fka \rpar}{\vartheta_{3} \lpar 0 | 2 \rmi \fka \rpar} \; . \nn
\err
The sum (instead of an integral) over all integer values of angular momentum in property (iii) asserts that the quantum phase-space for
angular coordinates is not equivalent to the classical one \cite{r7}. In order to prove this equality, one needs only to decompose the coherent states in either the angle or angular momentum basis and properly identify the realisation of Dirac or Kroenecker deltas, observing the periodicity of the angle variable or the infinite range of the angular momentum. Moreover, property (iv), which can be directly (but tediously) obtained, presents a perfect analogy with the
Cartesian case, where the $\vartheta_{3}$-function plays in this context the role that is reserved to the Gaussian function in the
linear case. In fact, the angle-angular momentum coherent states here constructed present a complete analogy (apart, at least, from a
phase factor) with that ones previously introduced in \cite{r10,r13,r14,r15,r16}.

\subsection{Uncertainty relations}
\begin{figure}[!th]
\centering
\begin{minipage}[b]{0.5\linewidth}
\includegraphics[width=\textwidth]{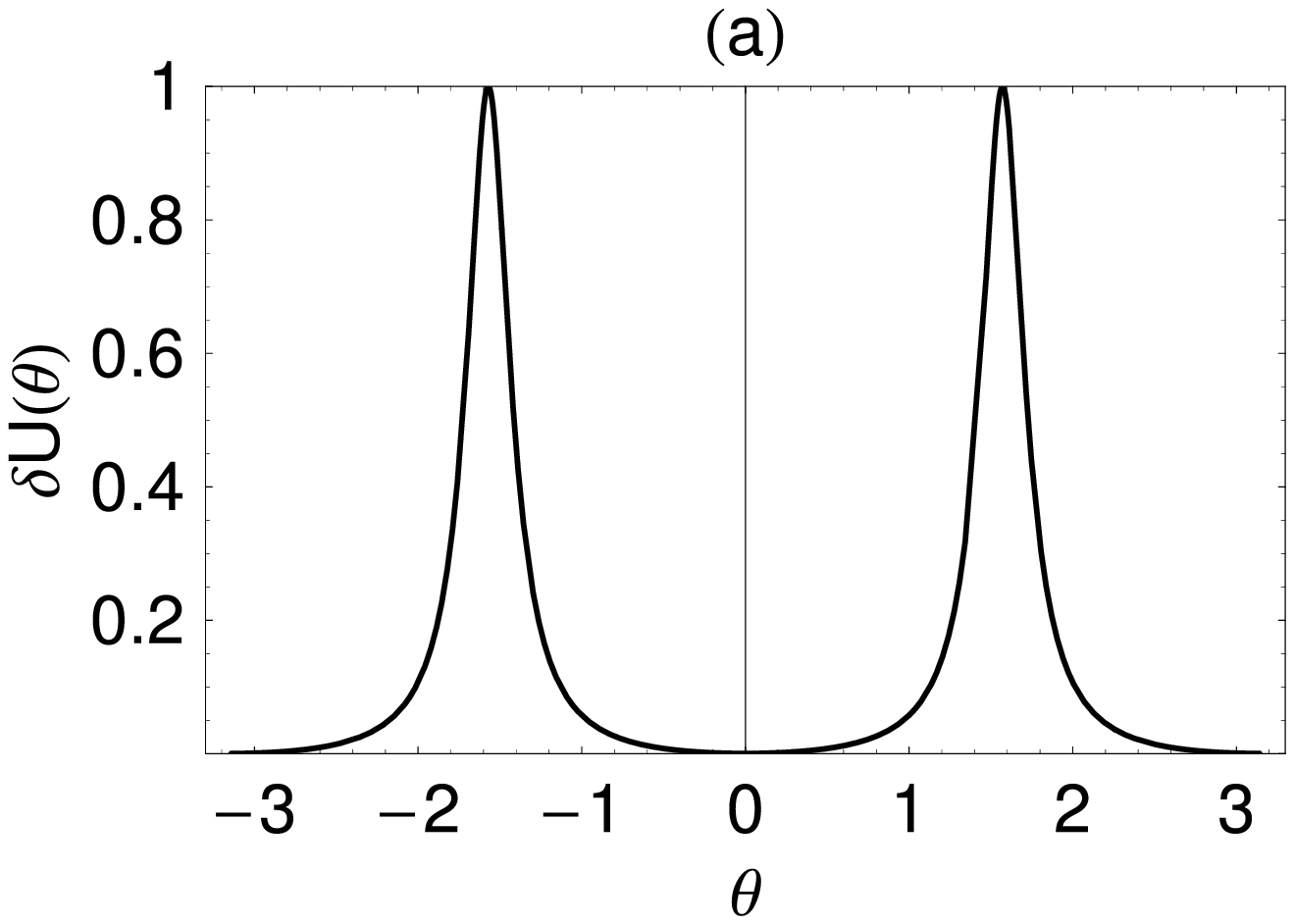}
\end{minipage} \hfill
\begin{minipage}[b]{0.5\linewidth}
\includegraphics[width=\textwidth]{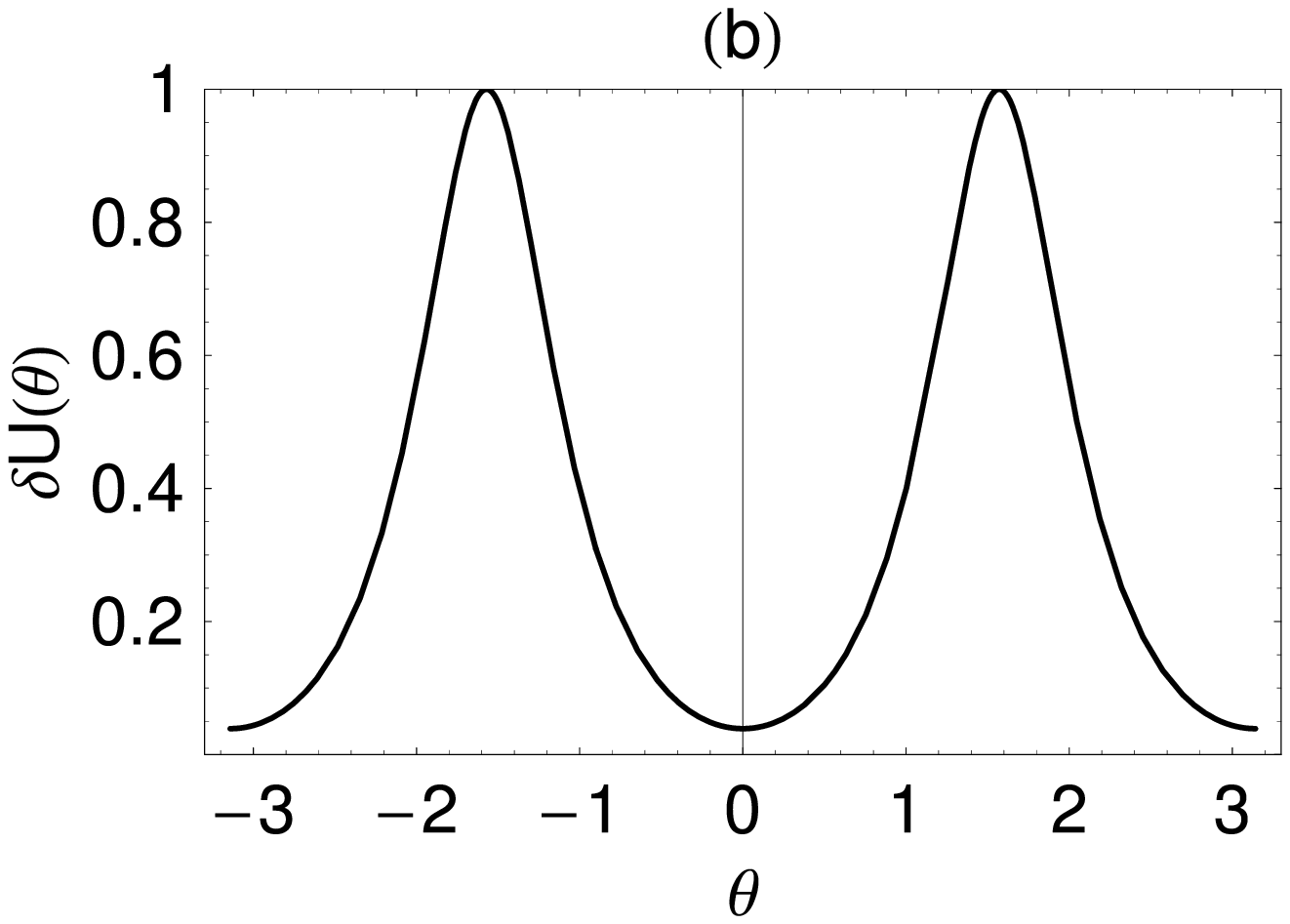}
\end{minipage} \hfill
\begin{minipage}[b]{0.5\linewidth}
\includegraphics[width=\textwidth]{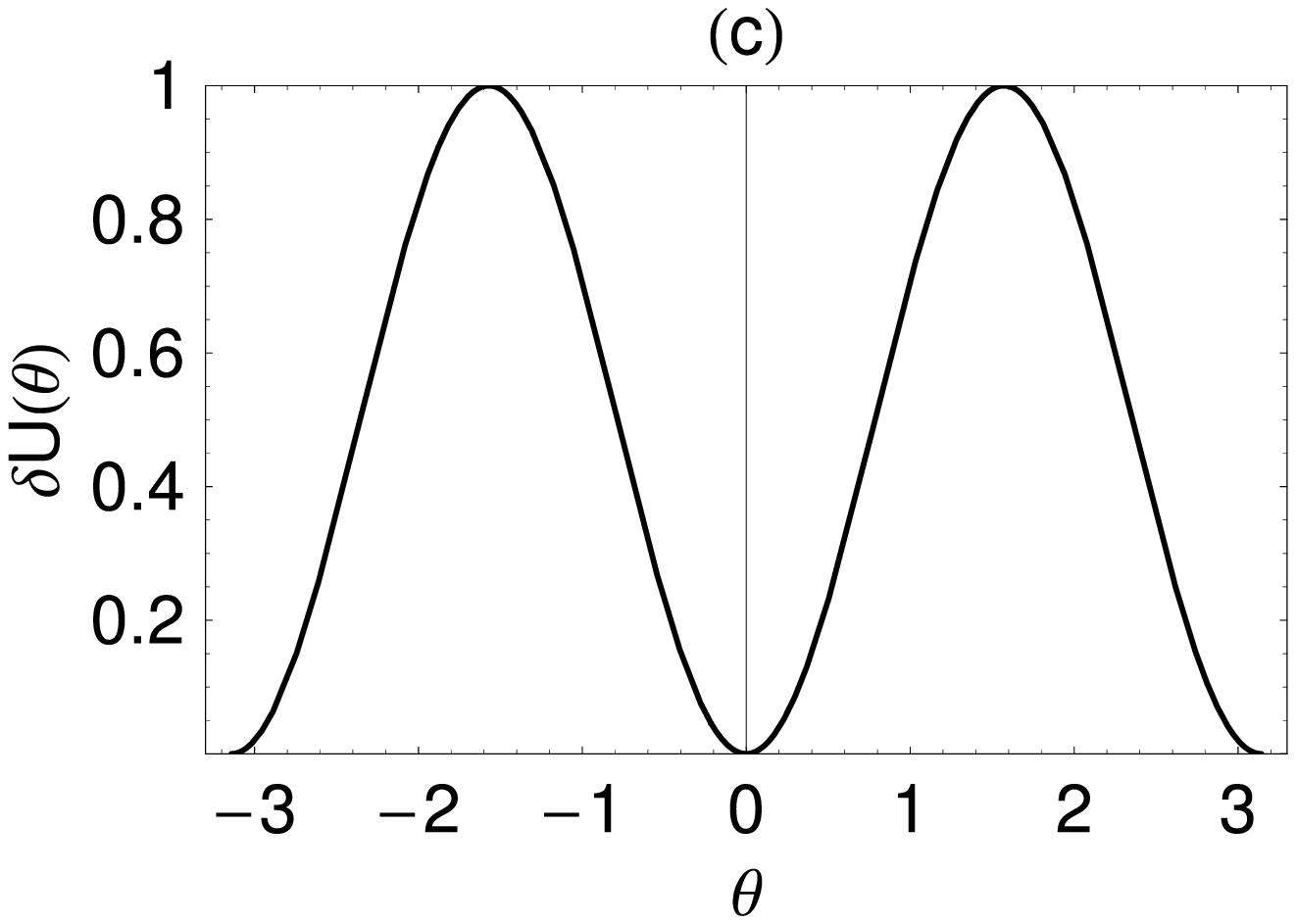}
\end{minipage}
\caption{Plots of $\delta \textrm{U}(\th)$ versus $\th \in [ - \pi,\pi )$ for (a) $\fka = 1/20 \pi$, (b) $\fka = 1/2 \pi$, and (c)
$\fka = 10/2 \pi$. These pictures show that $0 \leq \delta \textrm{U}(\th) \leq 1$, and for $\delta \textrm{U}(\th) = 0$ the coherent
states $\{ | m,\th \rg \}$ can be considered minimum uncertainty states. It is worth mentioning that the main differences between
intelligent and minimum uncertainty states have been discussed in \cite{r22} for the experimental context.}
\end{figure}
\hp To avoid some mathematical inconsistencies in the uncertainty relations following the commutation relation between the ${\bf J}$ and $\mathbf{\Theta}$ operators,
Carruthers and Nieto \cite{r18}, following Louisell \cite{r21}, introduced 
$\sin (\mathbf{\Theta})$ and $\cos (\mathbf{\Theta})$ which inherently embodies the periodicity property. For parity reasons, the one
appropriate to be concerned here is the non-symmetrycal relation
\be
\lb{uncert}
\textrm{U} \equiv \lg \Delta {\bf J} \rg^{2} \lg \Delta \sin (\mathbf{\Theta}) \rg^{2} \geq \frac{1}{4} \lg \cos (\mathbf{\Theta})
\rg^{2} \; ,
\ee
where the variances are explicitly evaluated through the relation $\lg \Delta {\bf O} \rg^{2} \equiv \lg {\bf O}^{2} \rg - \lg {\bf O}
\rg^{2}$. Now, let us consider the coherent states (\ref{dcs}) and their algebraic properties into this context. For instance, it is
straightforward to show that (\ref{uncert}) does not depend on the angular-momentum label, namely $\textrm{U} \equiv \textrm{U}(\th)$ (the symmetrical inequality given by Carruthers and Nieto is seen to be independent of both variables).
Besides, through the auxiliary relation
\bd
\delta \textrm{U}(\th) \equiv \frac{\lg \Delta {\bf J} \rg^{2} \lg \Delta \sin (\mathbf{\Theta}) \rg^{2} - (1/4) \lg \cos
(\mathbf{\Theta}) \rg^{2}}{\lg \Delta {\bf J} \rg^{2} \lg \Delta \sin (\mathbf{\Theta}) \rg^{2}} \; ,
\ed
it is possible to verify that for different values of $\fka$ (width of the $\vartheta_{3}$-function associated with the normalized
reference state) and $\th$ the angle-angular momentum coherent states are minimum uncertainty states. Figure 1 shows the plots of
$\delta \textrm{U}(\th)$ versus $\th$ for (a) $\fka = 1/20 \pi$, (b) $\fka = 1/2 \pi$ (value adopted in this work), and (c) $\fka = 10/2
\pi$. Note that $\delta \textrm{U}(\th)$ reaches its maximum value at the points $\th = \pm \pi/2$ in all pictures, while its minimum
value happens at the points $\th = 0, \pm \pi$. Since minimum uncertainty states are characterized by the mathematical condition
$\delta \textrm{U}(\th) = 0$, we can perceive that (b) shows in this case a small deviation around $4\%$ for the points located in
$\th = 0, \pm \pi$. Consequently, the right choice of parameters $\fka$ and $\th$ leaves relation (\ref{uncert}) arbitrarily close to
the equality (for similar results, see \cite{r10}).

\section{Mapping kernel}

\hp Following the mathematical procedure established in \cite{r23} (where the exposition, although in another context, is more detailed and instructive), given a set of coherent states it is always possible to define the
mapping kernel
\brr
\lb{kernel}
{\bf T}^{(s)}(m,\th) &\equiv& \sum_{l \in \mathbb{Z}} \int_{-\pi}^{\pi} d \alpha \exp \lbk - \rmi l (\th - \mathbf{\Theta)} \rbk
\exp \lbk \rmi \alpha (m - {\bf J}) \rbk \nn \\
& & \times \exp \lpar - \frac{\rmi}{2} \, l \alpha \rpar \lbk \mathcal{K}(l,\alpha) \rbk^{-s} \; ,
\err
where $\mathcal{K}(l,\alpha) \equiv \lg 0,0 | l,\alpha \rg$ denotes a particular overlap of coherent states explicitly calculated in
(iv), and $s$ represents a complex parameter satisfying $|s| \leq 1$. It is easy to show that the properties
\brr
&\mbox{(v)}& \; \Tr [ {\bf T}^{(s)}(m,\th) ] = 1 \; , \nn \\
&\mbox{(vi)}& \; \Tr[ {\bf T}^{(-s)}(m^{\prime},\th^{\prime}) {\bf T}^{(s)}(m,\th) ] = 2 \pi \, \delta_{m^{\prime},m} \,
\delta (\th^{\prime} - \th) \nn
\err
are promptly verified where, in particular, the last equality has been attained with the help of the auxiliary relation
\brr
\lb{trgeral}
\fl \qquad \Tr[ {\bf T}^{(s^{\prime})}(m^{\prime},\th^{\prime}) {\bf T}^{(s)}(m,\th)] &=& \sum_{l \in \mathbb{Z}} \int_{-\pi}^{\pi}
d \alpha \, \exp \lbr \rmi \lbk l (\th^{\prime}-\th) - \alpha (m^{\prime}-m) \rbk \rbr \nn \\
\fl & & \qquad \qquad \times \lbk \mathcal{K}(l,\alpha) \rbk^{-(s^{\prime} + s)} \; .
\err
Indeed, property (vi) guarantees that the decomposition of any bounded operator ${\bf O}$ in this basis assumes the form
\be
\lb{decomp}
{\bf O} = \sum_{m \in \mathbb{Z}} \int_{-\pi}^{\pi} \frac{d \th}{2 \pi} \, \mathcal{O}^{(-s)}(m,\th) {\bf T}^{(s)}(m,\th) \; ,
\ee
where the coefficients
\bd
\mathcal{O}^{(-s)}(m,\th) \equiv \Tr \lbk {\bf T}^{(-s)}(m,\th) {\bf O} \rbk
\ed
correspond to a one-to-one mapping between operators and functions belonging to a phase space characterized by the angle-angular
momentum variables. In addition, the mean value
\be
\lb{mean}
\lg {\bf O} \rg \equiv \Tr ( {\bf O} \ro ) = \sum_{m \in \mathbb{Z}} \int_{-\pi}^{\pi} \frac{d \th}{2 \pi} \, \mathcal{O}^{(-s)}(m,\th)
\, \mathcal{F}^{(s)}(m,\th)
\ee
can also be obtained from this decomposition, the parametrized function $\mathcal{F}^{(s)}(m,\th)$ being defined as the expectation
value of the mapping kernel (\ref{kernel}), i.e.,
\be
\label{qp}
\mathcal{F}^{(s)}(m,\th) \equiv \Tr \lbk {\bf T}^{(s)}(m,\th) \ro \rbk
\ee
with $\ro$ representing the density operator which describes an arbitrary physical system. As it will be seen, for $s=-1,0,+1$ the
parametrized function is directly related to the Husimi, Wigner, and Glauber-Sudarshan functions, respectively.

\subsection{Quasiprobability distribution functions}

\hp Now, let us consider the expansion for the projector of angle-angular momentum coherent states in the basis $\{ {\bf D}(l,\alf) \}$
as follows:
\be
\lb{proj}
| m,\th \rg \lg m,\th | = \sum_{l \in \mathbb{Z}} \int_{-\pi}^{\pi} \frac{d \alpha}{2 \pi} \, {\bf D}(l,\alf) \Tr \lbk {\bf D}^{\dagger}
(l,\alf) | m,\th \rg \lg m,\th | \rbk
\ee
where the Weyl algebra, obeyed by the displacement operators, ensures that
\be
\lb{trproj}
\Tr \lbk {\bf D}^{\dagger}(l,\alf) | m,\th \rg \lg m,\th | \rbk = \exp \lbk - \rmi ( l \th - m \alf) \rbk \mathcal{K}(l,\alf) \;.
\ee
Now, using the above result for the trace in Eq.(\ref{proj}), one immediately realizes that the obtained expression is exactly the same as the one of  Eq.(\ref{kernel}) for the particular case $s=-1$, which allows for the conclusion that
\be
| m,\th \rg \lg m,\th | ={\bf T}^{(-1)}(m,\th),
\ee
which perhaps is the central result of this paper. A lot of consequences follow from this fact. First, the Husimi function can be defined
within this context by means of a trace operation,
\be
\lb{husimi}
\fl \qquad \mathcal{H}(m,\th) \equiv \mathcal{F}^{(-1)}(m,\th) = \Tr \lbk | m,\th \rg \lg m,\th | \ro \rbk = \lg m,\th | \ro | m,\th
\rg \; .
\ee
On the other hand, if one considers $s=-1$ and ${\bf O} = \ro$ in equation (\ref{decomp}), we obtain the diagonal representation
\be
\lb{glauber}
\ro = \sum_{m \in \mathbb{Z}} \int_{-\pi}^{\pi} \frac{d \theta}{2 \pi} \, \mathcal{P}(m,\th) | m,\th \rg \lg m,\th | \; ,
\ee
where $\mathcal{P}(m,\th) \equiv \mathcal{F}^{(1)}(m,\th)$ plays the role of the Glauber-Sudarshan function.

In particular, for $s=0$ the mapping kernel reduces to the form already studied in references \cite{r7,r8,r9}. Hence, $\mathcal{F}^{(0)}
(m,\th) \equiv \mathcal{W}(m,\th)$ is nothing else than the Wigner function associated with the angle-angular momentum representation.

\subsection{Hierarchical structure}

\hp Next, we will obtain a hierarchical structure relating the Glauber-Sudarshan, Wigner, and Husimi quasiprobability distribution
functions. For this purpose, let us initially decompose the element ${\bf T}^{(s)}(m,\th)$ with the help of equation (\ref{decomp}),
\be
\label{tr2t}
\fl \qquad {\bf T}^{(s)}(m,\th) = \sum_{m^{\prime} \in \mathbb{Z}} \int_{- \pi}^{\pi} \frac{d \th^{\prime}}{2 \pi} \Tr \lbk
{\bf T}^{(- s^{\prime})}(m^{\prime},\th^{\prime}) {\bf T}^{(s)}(m,\th) \rbk {\bf T}^{(s^{\prime})}(m^{\prime},\th^{\prime}) \; .
\ee
The expression for $\Tr \lbk {\bf T}^{(-s^{\prime})}(m^{\prime},\th^{\prime}) {\bf T}^{(s)}(m,\th) \rbk \equiv
\mathfrak{Z}^{(s^{\prime}-s)}(m^{\prime}-m,\th^{\prime}-\th)$ is explicitly shown in equation (\ref{trgeral}). An immediate consequence
of this result is the link between different parametrized functions
\be
\lb{parfunc}
\fl \qquad \mathcal{F}^{(s)}(m,\th) = \sum_{m^{\prime} \in \mathbb{Z}} \int_{- \pi}^{\pi} \frac{d \th^{\prime}}{2 \pi} \,
\mathfrak{Z}^{(s^{\prime}-s)}(m^{\prime}-m,\th^{\prime} - \th) \, \mathcal{F}^{(s^{\prime})}(m^{\prime},\th^{\prime}) \; ,
\ee
where the term $\mathfrak{Z}^{(s^{\prime}-s)}(m^{\prime}-m,\th^{\prime}-\th)$ plays an important role in this process. Note that the
double Fourier transform in the right-hand side of equation (\ref{trgeral}) is well defined only for $\re(s^{\prime}-s) \geq 0$;
otherwise, the infinite summation for the angular momentum variable gives a divergent result. This implies in a hierarchical structure
analogous to that observed for the Cartesian case (we remark that this is not observed in the extended CG formalism for
finite-dimensional spaces \cite{r23}).

Two important results can be promptly reached through equation (\ref{parfunc}) for specific values of the complex parameters $s$ and
$s^{\prime}$, i.e.,
\brr
\lb{wgs}
\mathcal{W}(m,\th) &=& \sum_{m^{\prime} \in \mathbb{Z}} \int_{- \pi}^{\pi} \frac{d \th^{\prime}}{2 \pi} \, \mathfrak{Z}^{(1)}
(m^{\prime}-m,\th^{\prime}-\th) \mathcal{P}(m^{\prime},\th^{\prime}) \; , \\
\lb{huswig}
\mathcal{H}(m,\th) &=& \sum_{m^{\prime} \in \mathbb{Z}} \int_{- \pi}^{\pi} \frac{d \th^{\prime}}{2 \pi} \, \mathfrak{Z}^{(1)}
(m^{\prime}-m,\th^{\prime}-\th) \mathcal{W}(m^{\prime},\th^{\prime}) \; .
\err
Therefore, equations (\ref{wgs}) and (\ref{huswig}) exhibit a sequential smoothing which characterizes a hierarchical process among the
quasiprobability distribution functions in the angle-angular momentum phase space, $\mathcal{P}(m,\th) \rightarrow \mathcal{W}(m,\th)
\rightarrow \mathcal{H}(m,\th)$. Here,
\bd
\fl \qquad \mathfrak{Z}^{(1)}(m^{\prime}-m,\th^{\prime}-\th) \equiv \Tr \lbk {\bf T}^{(-1)}(m^{\prime},\th^{\prime}) {\bf T}^{(0)}
(m,\th) \rbk = \lg m^{\prime},\th^{\prime} | {\bf T}^{(0)}(m,\th) | m^{\prime},\th^{\prime} \rg
\ed
can be interpreted as a Wigner function evaluated for the angle-angular momentum coherent states labeled by $m^{\prime}$ and
$\th^{\prime}$. It is worth mentioning that
\be
\lb{husgs}
\mathcal{H}(m,\th) = \sum_{m^{\prime} \in \mathbb{Z}} \int_{- \pi}^{\pi} \frac{d \th^{\prime}}{2 \pi} \, | \lg m^{\prime},\th^{\prime}
| m,\th \rg |^{2} \, \mathcal{P}(m^{\prime},\th^{\prime})
\ee
establishes an additional relation which allows us to connect both the Husimi and Glauber-Sudarshan functions without the intermediate
process given by the Wigner function, with $| \lg m^{\prime},\th^{\prime} | m,\th \rg |^{2} = | \mathcal{K}(m^{\prime}-m,\th^{\prime} -
\th) |^{2}$ being the overlap probability for coherent states.

\section{Conclusions}

\hp One of the interesting features of the original CG approach is that it gives a clear cut answer to the problem of ordered expansions
in boson amplitude operators. As the own creation and annihilation operators in the present context are not a closed matter, we leave
this question to be discussed elsewhere, as our results might provide an useful approach to this problem.

In a more pragmatical sense, the above discussed quasiprobability distribution functions can be seen to be tailored for use in
treating the rotational degree of freedom in deformed physical systems, and discussing their semiclassical limits. In particular, some
previous attempts of introducing rotational coherent states have been put forth in the past whose aim was to treat the dynamics of
two-dimensional deformed systems in molecular physics \cite{r24}. In this connection, the use of the here proposed distributions in such
studies of rigid deformed systems dynamics, in the context of von Neunmann-Liouville formalism, seems to be a promising perspective.

Our results also seem to be quite suitable to deal with the problem of quantum rings, where a single electron can be trapped in a region
whose topology is exactly the one here regarded \cite{r25}. It is reasonable to expect that the quasiprobability functions might provide
convenient tools to obtain physical information from such systems.

\ack

\hp MR is supported by Funda\c{c}\~{a}o de Amparo \`{a} Pesquisa do Estado de S\~{a}o Paulo (FAPESP), Brazil, Project number
03/13488-0. Both MAM and DG (partially) are supported by Conselho Nacional de Desenvolvimento Cient\'{\i}fico e Tecnol\'ogico (CNPq),
Brazil. ECS is supported by Coordena\c{c}\~{a}o de Aperfei\c{c}oamento de Pessoal de N\'{\i}vel Superior (CAPES), Brazil.

\appendix

\section*{Appendix: Quantum Mechanics of the angle/angular momentum pair}
\setcounter{section}{1}

Let us initially consider a vector space spanned by a infinite
set of states $\left\{ |m\rangle \right\} _{m\in \mathbb{Z}}$ obeying
\begin{equation}
\mathbf{J}|m\rangle =m|m\rangle ,  \label{a1}
\end{equation}
where we take full advantage of the abstract Dirac notation. Here, for briefness, we deal only with the boson case. This set of
states is thus orthogonal, $\langle m^{\prime }|m\rangle =\delta _{m^{\prime
},m},$ and complete by assumption,
\[
\sum_{m=-\infty }^{\infty }|m\rangle \langle m|=\mathbf{1.}
\]
We then introduce a new family of states, constructed making use of the
Fourier coefficients 
\[
|\theta \rangle =\frac{1}{\sqrt{2\pi }}\sum_{m=-\infty }^{\infty }\exp (\rmi %
m\theta )|m\rangle ,
\]
where the label $\theta $ is a real number and, by construction, $|\theta
\rangle =|\theta +2\pi \rangle $ (therefore, one can always work within the
interval $-\pi \leq \theta <\pi $). It is possible to verify that these
states are orthogonal and complete,
\begin{eqnarray*}
\langle \theta ^{\prime }|\theta \rangle  &=&\delta ^{\lbrack 2\pi ]}(\theta
^{\prime }-\theta ),
\end{eqnarray*}
\begin{eqnarray*}
\int_{-\pi }^{\pi }d\theta |\theta \rangle \langle \theta | &=&\mathbf{1,}
\end{eqnarray*}
where the superscript $2\pi $ in the Dirac delta denotes that it is
different from zero whenever $\theta ^{\prime }-\theta =0\ ($mod $2\pi )$
(or, in a somewhat clumsier notation, $\delta ^{\lbrack 2\pi ]}(\theta
^{\prime }-\theta )=\sum_{k=-\infty }^{\infty }\delta (\theta ^{\prime
}-\theta +2\pi k)$). 

One may define now an angle operator by means of a spectral decomposition
\[
\mathbf{\Theta =}\int_{-\pi }^{\pi }d\theta \ \theta |\theta \rangle \langle
\theta |,
\]
and observe that the exponentials of these operators act as 
\begin{eqnarray}
\exp (-\rmi \theta \mathbf{J})|\theta ^{\prime }\rangle  &=&|\theta ^{\prime
}+\theta \rangle  \\
\exp (\rmi m\mathbf{\Theta })|m^{\prime }\rangle  &=&|m^{\prime }+m\rangle 
\end{eqnarray}
From the above results follows that Weyl algebra is observed, 
\begin{equation}
\label{weyl}
\exp (\rmi\theta \mathbf{J})\exp (\rmi m\mathbf{\Theta })=\exp (\rmi m\theta
)\exp (\rmi m\mathbf{\Theta })\exp (\rmi\theta \mathbf{J})\;.
\end{equation}
Therefore, these operators are the displacement generators in each other's
set of eigenstates -- which are connected through a Fourier transform--, and
it is exactly in this sense that one can say that they are canonically
conjugated. 

To see a typical realization of the action of the angular momentum operator,
if one considers an arbitrary state $|\psi \rangle ,$%
\[
|\psi \rangle =\int_{-\pi }^{\pi }d\theta \psi (\theta )|\theta \rangle ,
\]
where $\psi (\theta )$ is an arbitrary periodic function continuous in the
interval $[-\pi ,\pi ),$ the state $\mathbf{J}|\psi \rangle $ is seen to be
represented by 
\[
\mathbf{J}|\psi \rangle =\int_{-\pi }^{\pi }d\theta \psi (\theta )\mathbf{J}%
|\theta \rangle 
\]
and convenient use of the above resolutions of unity leads to 
\[
\mathbf{J}|\psi \rangle =\int_{-\pi }^{\pi }d\theta ^{\prime }\left[
\int_{-\pi }^{\pi }d\theta \psi (\theta )\frac{1}{2\pi }\sum_{m=-\infty
}^{\infty }m\exp (\rmi m(\theta -\theta ^{\prime }))\right] |\theta ^{\prime
}\rangle .
\]
Calling $I$ the term inside the square brackets one sees that it is in fact 
\[
\fl
I=-i\int_{-\pi }^{\pi }d\theta \psi (\theta )\frac{d}{d\theta }\frac{1}{2\pi 
}\sum_{m=-\infty }^{\infty }\exp (\rmi m(\theta -\theta ^{\prime
}))=-i\int_{-\pi }^{\pi }d\theta \psi (\theta )\frac{d}{d\theta }\delta
^{\lbrack 2\pi ]}(\theta -\theta ^{\prime }),
\]
as $\frac{1}{2\pi }\sum_{m=-\infty }^{\infty }\exp [\rmi m(\theta -\theta
^{\prime })]$ is a representation of the modulo $2\pi $ Dirac delta.
Integration by parts (and the periodicity of $\psi (\theta )$) then leads to 
\[
I=i\frac{d\psi (\theta ^{\prime })}{d\theta ^{\prime }},
\]
which means that 
\[
\mathbf{J}|\psi \rangle =\int_{-\pi }^{\pi }d\theta ^{\prime }\left[ i\frac{%
d\psi (\theta ^{\prime })}{d\theta ^{\prime }}\right] |\theta ^{\prime
}\rangle ,
\]
and thus $\mathbf{J}$ is seen as a derivative operator in the angle
representation, once one is dealing with states $|\psi \rangle $ constructed
out of periodic functions. One then can, in principle, look for the
commutator $\left[ \mathbf{J,\Theta }\right] ,$ although it is no trivial
matter to obtain an uncertainty relation from it, as it was discussed in \cite{r18}. We remark that, however,
the well defined Weyl commutation, obeyed by the displacement operators, is
all the algebra necessary for the purposes of this paper.


\end{document}